# Tri-state Buffer: Magnetic Vortex Transistor Based Logic Devices

**Authors:** Sucheta Mondal[1], Saswati Barman[2] and Anjan Barman[1,*]

[1]Department of Condensed Matter Physics and Material Sciences, S. N. Bose National Centre for Basic Sciences, Block JD, Sector III, Salt Lake, Kolkata 700 106, India

[2]Institute of Engineering and Management, Sector V, Salt Lake, Kolkata 700 091, India

*abarman@bose.res.in

**Abstract:** Magnetic analogue of electronic gates are advantageous in many ways. There is no electron leakage, higher switching speed and more energy saving in a magnetic logic device compared to a semiconductor one. Recently, we proposed a magnetic vortex transistor and fan-out devices based on carefully coupled magnetic vortices in isolated nanomagnetic disks. Here, we demonstrate a new type of magnetic logic gate based upon asymmetric vortex transistor by using micromagnetic simulation. Depending upon two main features (topology) of magnetic vortex, chirality and polarity, the network can behave like a tri-state buffer. Considering the asymmetric magnetic vortex transistor as a unit, the logic gate has been formed where two such transistors are placed parallel and another one is placed at the output. Magnetic energy given in the input transistors is transferred to the output transistor with giant amplification, due to the movement of antivortex solitons through the magnetic stray field. The loss and gain of energy at the output transistor can be controlled only by manipulating the polarities of the middle vortices in input transistors. Due to the asymmetric energy transfer of the antivortex solitons, we have shown successful fan-in operation in this topologically symmetric system. A tri-state buffer gate with fan-in of two transistors can be formed. This gate can be used as a 'Switch' to the logic circuit and it has technological importance for energy transfer to large scale vortex networks.

## I. INTRODUCTION:

Performance of modern computing devices relies on transport properties of electrons. The components of integrated circuits (ICs) *i.e.* transistor, diode, logic gate, are driven by the electron-hole pair generation and recombination. However, difficulty arises in semiconductor based nanoscale devices for their significant electron leakage and energy loss. Thus, several attempts have been made to construct all-optical polariton transistor [1] and strain induced logic operation in multiferroic systems [2] for viable large scale integration. Still cascadability is difficult to achieve during the optimization of power coupled from output of one level to the input of the next level. Due to the non-volatile storage capacity, nanomagnets with different aspect ratio [3, 4] have already been exploited for construction of logic devices and simplest OR gate to universal NAND gate have been engineered. Experimental techniques as well as micromagnetic simulations are being used to study magnetic domain wall based logic gates [5, 6]. In addition to these, very recent problems in magnetic skyrmions have opened up new opportunities to establish binary logic operation, using the helicity of anti-skyrmion [7]. In this context magnetic vortex structures [8] have also attracted huge attention for their unique properties. With higher areal density and convenient patterning procedure, magnetic vortex structure provides the dynamic response in GHz frequency [9] regime. Since a magnetic vortex has in-plane chiral spin texture with out-of-plane polarized core [10], we have freedom of switching the chirality and polarity of vortex core and magnetic analogue of digital circuits can be designed. A new type of energy transfer mechanism has been proposed in a chain of physically isolated magnetic vortices [11-14], where core gyration in one vortex can be induced to the neighboring vortices via magnetostatic coupling. Earlier, researchers have numerically demonstrated the logic operation and fan-out in the vortex network but no amplification of the signal was obtained [15]. More recently, the idea of magnetic vortex based transistor operation (MVT) [16] has been surfaced by D. Kumar *et al.* followed by asymmetric magnetic vortex transistor (AMVT) [17] with enhanced gain in the output, proposed by S. Barman *et. al.* with a successful fan-out operation. Besides those, H. Jung *et. al.* have experimentally shown that linearly placed three magnetic vortices can construct XOR and OR gate [18]. Further analysis shows that in a chain of three vortices, current induced gyration in the core of two side vortices can induce gyration in the middle vortex core by the robust mechanism of negligible-loss energy transfer via antivortex solitons [19]. N. Hasegawa also demonstrated by electrical excitation and detection technique an anisotropic energy transfer in magnetostatically coupled vortex pair [20]. However, the problem of reasonable amount of energy transfer to the side branches of a network during cascading remains unresolved.

A buffer with an active control input serves the purpose of amplified power transfer to large circuit and it also can act as a 'Switch' to the rest of the circuit. If this gate is active in a circuit, it will amplify the input of each cascading unit compensating all possible decays during the transfer of energy through a large network of fan-out branches. On the other hand, if this gate is inactive then no significant power can be transferred to the output despite of the higher strength of principal input. Here we have proposed a magnetic vortex based tri-state buffer using micromagnetic simulation where fan-in operation has been wisely exploited to achieve giant gain at the output of the logic gate. Here, the buffer has high impedance (HZ) state apart from conventional high (1) or low (0) states at the output. The AMVT with inter-vortex separations of 10 nm and 100 nm along with core polarities (1,-1,-1), proposed in our previous work [17] is chosen to be one unit in the tri-state buffer system. To construct a tri-state buffer, an AMVT with polarity combinations (1, 1,-1) was placed in between the two AMVTs with polarity combinations (1, -1,-1) with optimized separation. The idea behind this optimization was to drive the antivortex solitons to symmetrically carry the energy from both the input AMVTs to the output AMVT. The upper branch will act as the principal input and the lower branch will act as a control input (or 'Enable' input). Subsequently, the core polarities of the middle vortices of two input AMVTs are switched in four different combinations as (1,-1), (-1,-1), (-1, 1) and (1, 1). When the core polarity of the middle vortex at the lower branch is -1 (*i.e.* 'ON' state of the control input) the output vortex provides us with either gain or loss, depending upon the 'ON' or 'OFF' state (1 or 0) of upper branch. This can serve as source or sink of power to the rest of the circuit in a large vortex based network. However, when the core polarity of the middle vortex at the lower branch is 1 (*i.e.* 'OFF' state of the control input) this gate will have HZ state with negligible gain, compelling the system not to be a source or sink in the circuit. Basically, it disconnects the gate from rest of the circuit and can be thought of as a 'Switch'. Therefore, the proposed gate acts as a tri-state buffer with fan-in of two AMVT branches. This gate is similar to that in electronic circuit where tri-state logic gate is implemented in registers, bus drivers, flip-flops and also internally in ICs.

## II. RESULTS AND DISCUSSIONS:

A disk of permalloy ($Ni_{80}Fe_{20}$; Py hereafter) having diameter 200 nm and thickness 40 nm is used to simulate single magnetic vortex using Object Oriented Micromagnetic Framework (OOMMF) [21, 22]. This vortex has in-plane spin texture with counter clockwise (CCW) chirality and out-of-plane polarized core with polarity either up (P = 1) or down (P = -1). The dimension of vortex core is ~ 20 nm (see supplemental material) [23]. Three such vortices placed in a row with inter-vortex separation of 10 nm and 100 nm form a coupled vortex system with asymmetric separation. If a rotating in-plane magnetic field is applied at the leftmost vortex at the resonant frequency 1.27 GHz, the core starts gyrating in a circular orbit. When the applied magnetic field has an amplitude of 1.5 mT, the motion of vortex core remains in linear regime and vortex core gyration is governed by Thiele's equation [24]. The in-plane components of magnetization oscillate with time and the oscillations get damped eventually. During this core gyration, energy is transferred to right most vortex via the middle vortex. The inherent mechanism for this energy transfer is the movement of antivortex solitons through the stray field [16, 17]. As a result, the cores of other two vortices start gyrating in the same frequency but with different phases. Eventually the system reaches to a steady state.

We investigate the time evolution of spatial average of x-component of magnetization, $<m_x>$ for each vortex, and its corresponding energy spectral density (ESD). We normalize the magnetization by dividing the x-component of magnetization $M_x$ by saturation magnetization $M_s$. Subsequently, we performed Fourier transform to obtain the required ESD. The difference of ESD of the rightmost vortex and the leftmost vortex is considered as gain of the system, $B_{output}(dB) = ESD_{output}(dB) - ESD_{input}(dB)$. This actually corresponds to the amount of energy transferred from input vortex to the output vortex or simply the amplification in core gyration amplitude. The gain of the system depends on the amplitude of excitation given at the input, inter-vortex separations, combination of core polarities and aspect ratio of the magnetic disk forming the vortex. This system is shown to behave as an asymmetric magnetic vortex transistor (AMVT) with enhanced gain 28.2 dB in 'ON' state, when the polarity combinations of the three vortices are (1,-1,-1) [17].

Here, we have used the above mentioned asymmetric magnetic vortex transistor (AMVT) as unit for the fan-in operation. As shown in Fig. 1 (a) two AMVTs with polarity combination (1,-1,-1) and CCW chirality are placed in parallel, with a separation of 200 nm ($S_y$) in y-direction and these have been used as the inputs of the logic gate. Here, the upper branch is acting as the principal input (A) and the lower branch is considered to be the 'Enable' input (EN). A third AMVT (Y) with polarity combination (1, 1,-1) is placed at the output symmetrically in y-direction, between the input branches, at a distance of 10 nm from the output vortices (O1 and O2) in the positive x-direction (see supplemental material for further details) [23]. The difference in ESDs between the input vortex (I3) and the output vortex (O3) of this chain is considered to be the gain at the output branch 'Y'.

Figure 1 (b) shows the schematic representation of the tri-state buffer, which is a conventional buffer with an extra input. In absence of any external bias voltage at I3, anti vortex solitons transfer the energy from input branches to the output through fan-in mechanism (see Supplemental Material for the movie) [23]. If we switch the core polarities of the middle vortices of input AMVTs (A and EN), gain in individual input branches will be modified and that will cause a noticeable change in the gain at the output vortex (O3). This gain may be positive or negative. We have considered positive and negative gain as 1 or 0 respectively, in case of logic operation. Basically, the magnitude of positive gain should be large enough in comparison to the negative gain to form a sensible logic gate.

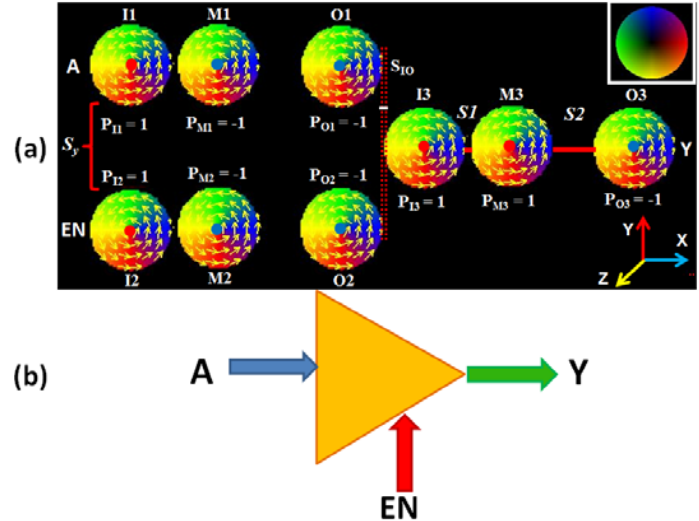

**Figure 1:** (a) The ground state spin configuration of magnetic vortex based tri-state logic gate using Py disk of diameter of 200 nm and thickness of 40 nm arranged in three branches with the separation between left and centre vortex ($S1$) of 10 nm and the separation between centre and right vortex ($S2$) of 100 nm in each branch. Distance between two input branches (A and EN) is $S_y$ = 200 nm. The output branch (Y) is placed at the middle of two branches with 10 nm separation in x-direction. (b) Schematic of conventional symbol representing a tri-state buffer.

We have optimized the distance between two parallel AMVTs (A and EN), aiming to achieve same gain in both the output ends (O1 and O2). This will help to maximize the power transfer from the input circuit to the output of the whole system, when both the input AMVTs will be in 'ON' state.

### A. Two parallel AMVTs:

If two AMVTs are placed in parallel configuration (as shown in Fig. 2 (a)), magnetostatic coupling between the unsaturated spin of the edges of vortices depends strongly on the distance between these branches. During vortex core gyration the dynamic stray field also gets modified according to this coupling and the core polarities also depend on this coupling. Earlier studies on the coupled vortex system have revealed the mechanism behind energy transfer between two vortices when they behave as coupled oscillator [16]. Depending upon the relative core polarities, one primary soliton tries to move and transfer energy from input vortex to the output. Since the present system consists of many vortices with different core polarities, the energy transfer mechanism will also be much more complicated. To explore the role of inter-branch separation upon the magnetostatic coupling and finally to achieve maximum gain in the output vortices of the two branches, the separation in y-direction ($S_y$) has been varied from 30 nm to 220 nm, at 10 nm interval.

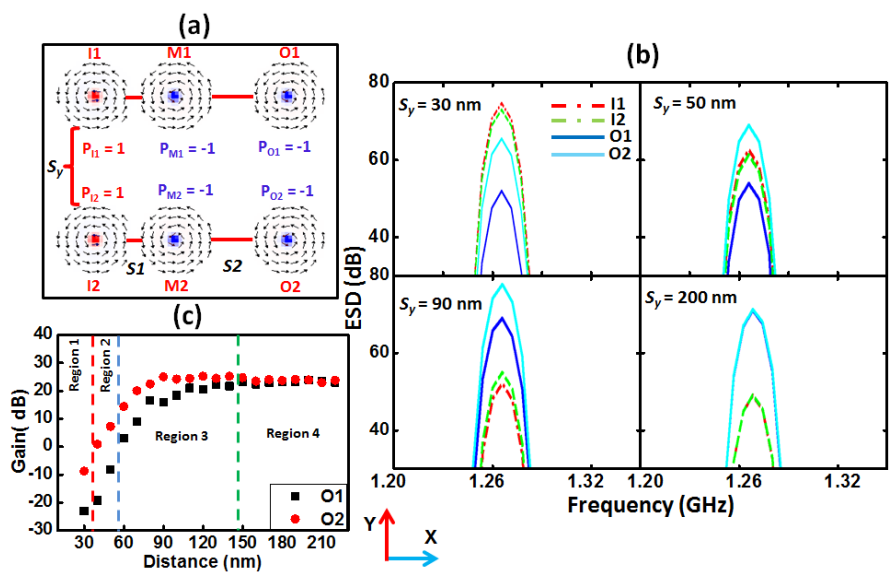

**Figure 2:** (a) The ground state spin configuration of two parallel placed conventional AMVTs (A and EN) with the separation in between them to be $S_y$ varied from 30 nm to 220 nm. (b) Energy spectral densities (ESDs) of the input and output vortices varying $S_y$ = 30, 50, 90 and 200 nm. (c) Variation of gain (B) with $S_y$ for this system.

We have found four different regions, according to combination of gains in the upper and lower AMVTs. Those are, (i) $S_y$ = 30 nm, (ii) 40 nm ≤ $S_y$ < 60 nm, (iii) 60 nm ≤ $S_y$ < 150 nm and (iv) 150 nm ≤ $S_y$ ≤ 220 nm. For $S_y$ less than 30 nm or greater than 220 nm, core switching takes place (See Fig. 1S of supplemental material)[23]. As the resonant frequency for single vortex has been found to be 1.27 GHz for coupled AMVTs the fundamental mode for vortex core gyration is also considered to be same. Figure 2 (b) shows that the gains at the output ends (O1 and O2) are -23.0 dB and -8.7 dB, respectively for $S_y$ = 30 nm. With increasing separation, the gains at both the ends increase gradually. For $S_y$ = 50 nm, output vortex of the lower branch (O2) experiences gain (7.2 dB) and the output vortex of the upper branch (O1) suffers loss (- 8.1 dB). For $S_y$ = 90 nm, both the vortices (O1 and O2) have gain but those are not equal (16.0 dB and 25.0 dB, respectively). This increase in gain continues with increase in separation ($S_y$) and for $S_y$ = 200 nm the gain is enhanced to 24.0 dB at both the output ends. Figure 2 (c) shows the variation of gain with separation ($S_y$) for both O1 and O2. Thus, reegion 1 corresponds to the separation range where both O1, O2 have loss. Region 2 corresponds to 40 nm ≤ $S_y$ < 60 nm, where O2 has gain but O1 has loss. This asymmetry continues for region 3 also, where for 60 nm ≤ $S_y$ < 150 nm, the O2 has preferential increment in gain O1. This variation in gain is not only due to monotonous increment of ESDs in the output vortices but also due to the variation of ESDs of the two input vortices. The variation of ESDs of these vortices with increasing separation is shown in Fig. 1S (b) and (c) of supplemental material, with more clarity [23]. But for region 4, *i.e.* 150 nm ≤ $S_y$ ≤ 220 nm, similar as well as maximum gain of 24.0 dB is achieved in both the output vortices (see Fig. 2S in the supplemental material for the quantification of core shift) [23]. In literature, the variation of in-plane component of magnetization ($m_x$) as well as the ESD, have been used as an indicator for the amplitude of core gyration [25-26]. To obtain deeper understanding of the mechanism behind the energy transfer by the anti vortex solitons, nature of dynamic stray field distribution simulated using a home-grown code [16], has been studied carefully. The detailed temporal evolutions of the magnetic stray fields are provided in supplementary movies (Movie1 to Movie 4).

Figure 3 shows the stray field distributions for magnetostatically coupled two AMVTs with four different inter-branch separations at time (t) = 40 ns, after the system reaches to steady state. For $S_y$ = 30 nm, one antivortex rotating in CCW direction, controls the dynamics of the stray fields near I1 and I2. The cores of two input vortices gyrate with greater amplitude than the output vortices. Between the middle and outermost vortices (M1, O1 and M2, O2), three antivortex solitons are constantly exchanging energies among themselves. This energy exchange results in feedback of energy from the output vortices to the input due to which significant drop in ESDs for O1 and O2 occurs. The negative gain in O1 and O2 are different because one antivortex soliton from the lower branch (EN) is carrying energy to the input vortex (I1) of the upper branch (A) by increasing its core gyration amplitude slightly. As the ESD of I1 increases, gain at O1 decreases ($B_{O1} < B_{O2}$). For $S_y$ = 50 nm, the antivortex solitons transfer energy in a different manner. In this case, interaction among the three solitons is weaker in comparison to $S_y$ = 30 nm. The antivortex soliton moving close to O1 appears and gets absorbed near the upper branch (A) only. Other two solitons move in circular paths in isolation. No feedback of energy from the output to the input vortices is present here. Another antivortex soliton tries to circulate energy around the outermost part of the circuit (near O1 and O2). This hinders the loss of energy in the output vortex of the lower branch. In addition, another antivortex soliton is transferring energy from the middle vortex of the lower branch (M2) to the input vortex of the upper branch (I1) providing sufficient core gyration amplitude resulting in loss at the upper branch (O1). For $S_y$ = 90 nm, the soliton gyrating in the leftmost part of the system keeps the dynamic stray field rotating between I1 and I2.

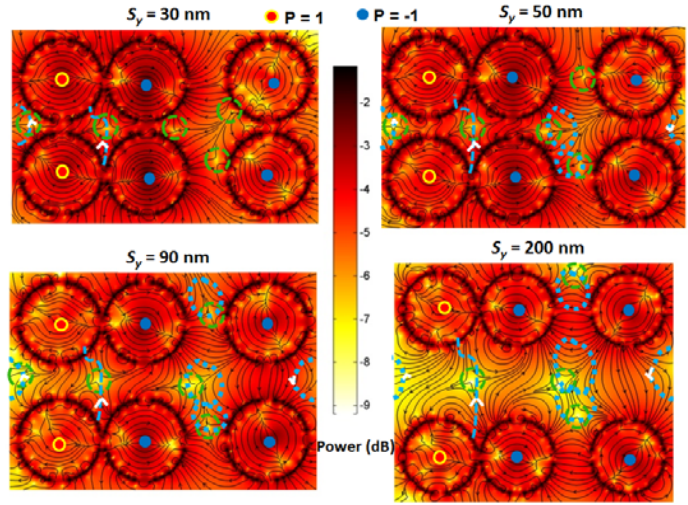

**Figure 3:** Stray field distributions for magnetostatically coupled two AMVTs with four different inter-disk separations at time, t = 40 ns, are shown. The green arrows represent the paths of the antivortex solitons and the antivortices are marked by green circles. Their movements are mapped by using blue dotted lines. The colour bars are shown at the middle of the figure. The contour coloring is based on the sum of squares of x- and y-components of the stray field and the color bar is in dB. The cores of the vortices are marked by red and blue dots corresponding to 1 and -1 polarities of the cores.

In this case, the separation between two vortices has been increased, so interaction with the antivortex is much less than that in the previous cases. In addition to that, one antivortex soliton is now rotating in a close loop near the M1 and O1 in the upper branch, instead of getting absorbed there. Other two antivortices gyrate freely, modifying the stray field near O2. Another antivortex soliton seems to distribute the energy in two outputs by gyrating in CW direction. A progress towards successful fan-in operation starts from here, because both the output ends achieve large amount of gain though the gains are significantly different.

Finally for $S_y$ = 200 nm (as shown in Fig. 3), the size of the antivortex gyrating at the leftmost end has decreased and the distance from both the input vortices has been increased, causing lesser gyration amplitude of the input cores. This time the cores of I1 and I2 gyrate in similar amplitude and phase. The antivortices gyrating at the middle of two branches are now covering larger path. Another soliton moving in between M1 and O1 is transferring more energy to O1. The soliton at the rightmost end is now rotating in a circular path causing uniform distribution of energy in both the output vortices (O1 and O2) and symmetric gains are obtained at O1 and O2. Therefore, we have finally achieved symmetric gains at both the output ends (O1 and O2) by systematically changing the inter-branch separation between two AMVTs.

### B. **Construction of a tri-state buffer:**

We have placed another AMVT with same geometrical configuration and polarity combination (1, 1, -1) with a separation between the rightmost edges of the two input branches and the leftmost edge of the output branch ($S_{IO}$) of 10 nm as shown in Fig. 4 (a). Among the four regions shown in Fig. 2 (c), we have found region 4 to be most appropriate to construct a logic gate with high contrast between the amount of loss and gain at the output vortices and we have kept the separation along y-direction ($S_y$) between two input braches to be 200 nm. To initiate the gyrotropic vortex motion and to activate transistor operation at two input branches (A and EN), continuously rotating magnetic field of 1.5 mT amplitude has been applied which is analogous to the biasing voltage of BJT [16, 17]. The cores of I1 and I2 start gyrating and the movement of antivortex solitons in the corresponding stray field transfers the energy to the neighbouring vortices. Finally, after few tens of nanoseconds, the energy distribution process comes to equilibrium. All the cores gyrate at same frequency but different phases and amplitudes. The energy from input branches is transferred to the output AMVT via fan-in mechanism similar to electronic systems. As discussed earlier, the difference in ESDs between O3 and I3 is considered as the gain (loss) at the output end which will be described as 1 (0) states at the output for logic operation. If the core polarities of any or both of the middle vortices in the input branches (A and EN) are switched from -1 to 1, the input AMVTs will not anymore operate in the 'ON' state, but instead they will switch to 'OFF' state [17]. Figure 4 (b) shows that when both the core polarities of M1 and M2 are -1, then ESD at I3 and O3 are 10.0 and 61.0 dB, respectively. This causes enormous gain at the Y end ($B_{O3}$ = 51.0 dB). When the core polarity of M1 is switched to 1 and that of M2 remains unchanged, then the upper branch will act in the 'OFF' state and the lower branch will be in 'ON' state. We have observed a significant loss in the output ($B_{O3}$ = -19.0 dB), which indicates that the output branch (Y) is controlled primarily by the upper input branch (A). If the polarities of M1 and M2 are exchanged, then $B_{O3}$ = 13.0 dB, which is much smaller than the gain with $P_{M1}$ = -1, $P_{M2}$ = -1. If the core polarities of both M1 and M2 are 1, then the ESDs of I3 and O3 are almost same and $B_{O3}$ = 2.0 dB.

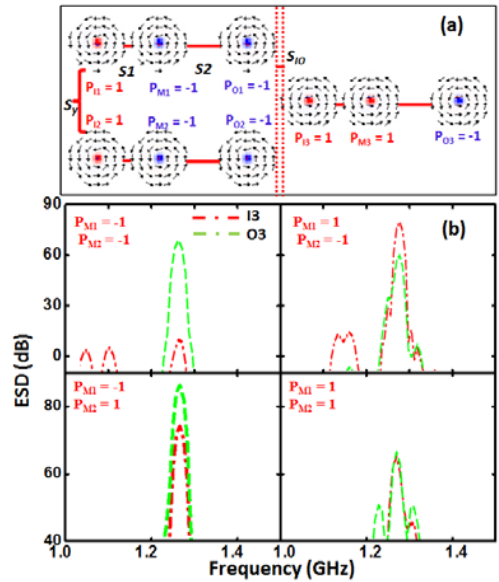

**Figure 4:** (a) The ground state spin configuration of two parally placed conventional AMVTs (A and EN) having the separation $S_y$ = 200 nm with another AMVT (1, 1, -1) placed symmetrically at 10 nm distance from the two branches. (b) Energy spectral densities (ESDs) of the input and output vortices (I3 and O3) varying polarity combinations of M1 and M2.

From the above findings, we understand that our system responds differently (logically) to core polarity change of upper (A) and lower (EN) input branches. As a result, this can act as an all-magnetic logic gate. It follows a logic operation, similar to the electronic tri-state buffer. This gate has an extra input called Enable input (EN). Depending upon the signal of EN, a new state at the output (Y) is formed. In conventional buffer, when input signal is high (1) or low (0) output shows the same. At these states, the gate actually works as source or sink to the rest of the output circuit. Energy can flow either to the circuit or can be absorbed. Tri-state logic has high impedance state (HZ) where energy can neither be supplied nor be absorbed in the circuit, when the extra input is not enabled (0). Therefore, this gate can be used as a 'Switch' which can isolate the input circuit from rest of the circuit in presence of an extra input (EN). The logic operation of this system will follow the following table:

**Table1**: (a) Logic operation in terms of polarity change of the vortex network is defined here. (b) For a Tri-state buffer the logic operation is represented in terms of binary numbers.

| (a) | $P_{M1}$ | $P_{M2}$ | $B_{O3}$ (dB) | State |
|---|---|---|---|---|
| | -1 (upper branch is in 'ON' state) | -1 (lower branch is in 'ON' state) | 51.0 | High |
| | 1 (upper branch is in 'OFF' state) | -1 (lower branch is in 'ON' state) | -19.0 | Low |
| | -1 (upper branch is in 'ON' state) | 1 (lower branch is in 'OFF' state) | 13.0 | Tri-state |
| | 1 (upper branch is in 'OFF' state) | 1 (lower branch is in 'OFF' state) | 2.0 | Tri-state |

| (b) | A | EN | Y | State |
|---|---|---|---|---|
| | 1 | 1 | 1 | act as source |
| | 0 | 1 | 0 | act as sink |
| | 1 | 0 | HZ | High Impedance |
| | 0 | 0 | HZ | High Impedance |

According to the core polarities (1 or -1) of M1 and M2, the convention used for input states are 0 and 1, respectively. The gain at O3 of the output AMVT ($B_{O3}$), is considered to be '1' and '0' when there is enormous gain or noticeable loss. However, negligible amount of gain found at the output end is not enough to transfer the energy significantly in extended networks and the corresponding state is the third state, i.e. a high impedance state. Therefore, this system has similarities with electronic tri-state logic gate and this can be a reliable candidate for all magnetic computation.

To gain more gain insight into the energy transfer mechanism in this fan-in based logic operation, we have systematically analyzed the temporal evolution of stray field distribution in Fig. 5. The detailed temporal evolutions of the magnetic stray fields are provided in supplementary movies (Movie 5 – Movie 8). When the polarity combinations of the two middle vortices in the upper and lower branches of the input AMVTs are (-1, -1), the gyration radius (gyration amplitude) of the core of I3 is much smaller than that of core of O3. The paths of the antivotices, distributing energy in the input branches are same as described earlier in Fig. 3, for $S_y$ = 200 nm. In addition to these, one antivortex near the output branch moves in almost a circular closed path as shown in Fig. 5(a). This soliton tries to circulate near M3 and O3 and transfers energy to the outermost vortex. While moving between O1, O2 and I3, the antivortex soliton gets absorbed in the I3 which affects the dynamics of the adjacent stray field. Consequently, the core of I3 now rotates with a smaller radius while the core of O3 continues to rotate as before providing almost 51.0 dB gain resulting in successful fan-in operation. As shown in Fig. 5 (b) when the polarity combinations of the two middle vortices in the upper and lower branches of the input AMVTs are (1, -1) the paths of the antivortex solitons change drastically in the input branches.

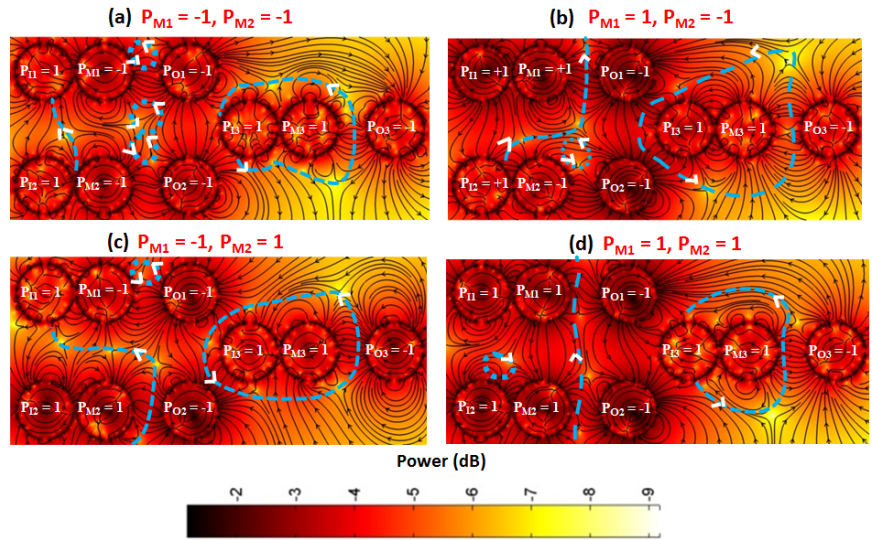

**Figure 5:** Stray field distributions for tri-state logic gate with inter-disk separations $S_y$ = 200 nm, and polarity combinations of (a) $P_{M1}$ = -1, $P_{M2}$ = -1, (b) $P_{M1}$ = 1, $P_{M2}$ = -1, (c) $P_{M1}$ = -1, $P_{M2}$ = 1 and (d) $P_{M1}$ = 1, $P_{M2}$ = 1 at time, t = 25 ns, are shown. The blue arrows represent the paths of the antivortex solitons. The colour bars are shown at the middle of the figure. The contour colouring is based on the sum of squares of x- and y-components of the stray field and the colour bar is in dB.

Since the direction of core gyration depends upon the core polarity, the direction of core gyration changes when the core polarity of M1 and M2 changes. Adjacent stray field gets arranged in such a way that antivortex solitons try to follow new path. Therefore, the gains in O1 and O2 are found to be quite different this time. Eventually the energy transfer procedure from input to output gets modified. The single soliton in the output branch moves between O1, O2 and I3 quite freely, giving rise to larger gyration radius to the core of I3.. However, when this soliton moves between M3 and O3, it stays in a meta-stable state for almost 0.3 ns. So the core of O3 gyrates with smaller amplitude (radius) and the corresponding power in O3 is 19.0 dB less than that in I3. Therefore, the system is not able to transfer energy any further to extended network. Figure 5 (c) shows the stray field distribution of the system when core polarities of M1 and M2 are -1 and 1, respectively. In this case, series of antivortices move from O2 to I1 (lower branch to upper branch) transferring energy and creating a feedback to the input ends. This reduces the $B_{O2}$, indicating that the lower branch AMVT is operating in 'OFF' state. The upper branch still operates in 'ON' state and fan-in operation continues. As a result, $B_{O3}$ is quite low (13.0 dB) and this will not be able to trigger a new circuit with reasonable power. When polarity combination of M1 and M2 are (1, 1)(as shown in Fig. 5 (d)), one antivortex is found to be gyrating at the input branch near I2, but another train of solitons passes through M2, O2 and M1, O1. This equalizes the energy distribution at upper and lower branches of the input circuit. So the cores of O1 and O2 gyrate in same phase resulting in very little variation in the stray fields in between O1, O2 and I3. In addition to that, a single soliton in the output branch gyrates in a similar path as in (a) and is absorbed in I3. Finally, the ESD of O3 is found to be almost same as I3, resulting only in 2.0 dB gain. So our analyses clearly show that $B_{O3}$ controlled by the upper AMVT branch (A), when the EN input is in 'ON' state. However, when EN is in 'OFF' state (polarity of middle vortex M2 is 1) $B_{O3}$ is suppressed, irrespective of the polarity of the M1. This is the 'high impedance' state (HZ) of the buffer as discussed earlier.

Using logic gates with higher fan-in will help reducing the depth of a logic circuit. However physical logic gates with a large fan-in tend to be slower than those with a small fan-in because of the complexity in the input circuit. Here the fan-in operation demonstrated in the all-magnetic network with two input branches is particularly interesting, because it not only transfers the energy continuously to the output branch but also the output can achieve giant gain when the system stabilizes over nanosecond timescale. Consequently, the construction of magnetic analogue of tri-state buffer (a three-state logic gate) is mainly driven by this fan-in mechanism where a high impedance (HZ) state can effectively remove the device's influence from the rest of the circuit. In case of electronic tri-state buffers, if more than one device is electrically connected to another device, putting an output into the HZ state is often used to prevent short circuits, or one device driving high (1) against another device driving low (0). Those can also be used to implement efficient multiplexers with large numbers of inputs, shared electronic bus etc. Similarly, magnetic analogue of the three-state logic can reduce the number of branches needed to drive a set of other components in an all-magnetic network.

## Conclusion:

In summary, we have constructed an all-magnetic tri-state buffer by using successful fan-in operation in AMVT for the first time and demonstrated the detailed mechanism behind construction of this tri-state buffer. To achieve an appreciable amount of gain and loss during construction of this tri-state buffer, we optimized the inter-branch separation from 30 nm to 220 nm of two parallel input AMVTs with core polarities (1,-1,-1). Subsequently, we found that the gains at both the output vortices in two input parallel AMVTs strongly depend upon the paths of the antivortices rotating between the two branches. In the four different regimes of inter-branch separations, the combinations of loss and gain at upper and lower branches are different. We choose the regime, where the gains are significant in the both the input AMVT branches and this allows significant energy transfer to the output AMVT for a successful fan-in operation. Subsequently, we construct the tri-state buffer by adding one output AMVT. The extra input is the enable input and the system behaves like a logic gate by following tri-state logic operation similar to the electronic tri-state buffer. The upper input branch acts as the principal input and lower input branch acts as an enable input (EN). The system has three states such as 'transmission' state, 'absorption' state and 'high impedance' state. When 'EN' is in 'ON' condition, the system can behave as either a source or a sink, depending upon the configuration of the principal input. Here the binary state output is similar to the state of principal input. In the 'high impedance' state, 'EN' is in 'OFF' condition and the system neither transmits energy nor absorbs energy. Therefore, this gate can also be used as a 'Switch' to isolate the input circuit from the rest of the circuit. Analysis of the complicated stray field dynamics, creation and absorption of antivortex solitons and their trajectories can explain the inherent mechanism of the logic operation. Optimization of the geometrical structure is very crucial to obtain the tri-state logic operation. Our findings can provide the platform for designing of magnetic vortex based logic gate and can finally lead towards all-magnetic computing devices.

Acknowledgement:

We acknowledge the financial support from the S. N. Bose National Centre for Basic Sciences (Grant No.: SNB/AB/18-19/211) and DST (Grant No.: DST/AB/14-15/133). S. M. acknowledges Department of Science and Technology, India for INSPIRE fellowship.